\documentclass{aa}

\usepackage{graphicx}
\usepackage{txfonts}
\usepackage{natbib}
\bibpunct{(}{)}{;}{a}{}{,}

\begin{document}

\title{Theoretical mass loss rates of cool main-sequence stars}

\author{V. Holzwarth\inst{1,2} \and M. Jardine\inst{2}}
\authorrunning{V. Holzwarth \& M. Jardine}
\offprints{V. Holzwarth} 

\institute{Max-Planck-Institut f\"ur Sonnensystemforschung,
Max-Planck-Strasse 2, 37191 Katlenburg-Lindau, Germany \\
\email{holzwarth@mps.mpg.de}
\and
School of Physics and Astronomy, University of St Andrews, 
North Haugh, St Andrews, Fife KY16 9SS, Scotland \\
\email{mmj@st-andrews.ac.uk}}

\date{Received ; accepted}

\abstract
{The stellar mass loss rate is important for the rotational evolution of
a star and for its interaction with the circumstellar environment.
The analysis of astrospheric absorption features enables an 
empirical determination of mass loss rates of cool stars other than the
Sun.}
{We develop a model for the wind properties of cool main-sequence
stars, which comprises their wind ram pressures, mass fluxes, and
terminal wind velocities.}
{The wind properties are determined through a polytropic magnetised
wind model, assuming power laws for the dependence of the thermal and
magnetic wind parameters on the stellar rotation rate.
We use the empirical data to constrain theoretical wind scenarios,
which are characterised by different rates of increase of the wind
temperature, wind density, and magnetic field strength.}
{Scenarios based on moderate rates of increase yield wind ram pressures
in agreement with most empirical constraints, but cannot account for
some moderately rotating targets, whose high apparent mass loss rates
are inconsistent with observed coronal X-ray and magnetic properties.
For fast magnetic rotators, the magneto-centrifugal driving of the
outflow can produce terminal wind velocities far in excess of the
surface escape velocity.
Disregarding this aspect in the analyses of wind ram pressures leads to
overestimations of stellar mass loss rates.
The predicted mass loss rates of cool main-sequence stars do not exceed
about ten times the solar value.}
{Our results are in contrast with previous investigations, which found
a strong increase of the stellar mass loss rates with the coronal X-ray
flux.
Owing to the weaker dependence, we expect the impact of stellar winds
on planetary atmospheres to be less severe and the detectability of
magnetospheric radio emission to be lower then previously suggested.
Considering the rotational evolution of a $1\,{\rm M_{\sun}}$ star, the
mass loss rates and the wind ram pressures are highest during the
pre-main sequence phase.}


\maketitle

\section{Introduction}
\label{intro}
Solar-like stars with hot coronae are expected to lose mass in the form
of stellar winds \citep{1960ApJ...132..821P}.
In contrast to the P Cygni line profiles of the massive winds of hot
stars and young T Tauri stars \citep[e.g.][]{2005ApJ...625L.131D}, the
tenuous and highly ionised outflows of cool stars yield no detectable
radiative signatures and cannot be diagnosed directly.
\citet{2002ApJ...574..412W} devised an indirect method to deduce mass
loss rates of cool stars from specific properties of the astrospheres
blown into the ambient interstellar medium by their stellar wind (see
also \citealt{1998ApJ...492..788W} and \citealt{2004LRSP....1....2W}).
Hot neutral hydrogen in the heliosphere, mainly in the form of a
\ion{H}{I} wall between the heliopause and the bow shock, leads to the
formation of a red-shifted absorption feature in stellar Ly$\alpha$
emission lines, whereas a corresponding structure in the astrosphere
around a star causes a specific blue-shifted feature.
The size and absorption capabilities of the astrosphere depend on the
strength of the stellar wind as well as on the properties (i.e.\
density, relative velocity) of the ambient interstellar medium (ISM).

By fitting synthetic absorption features, obtained from hydrodynamical
simulations of the wind-ISM interaction
\citep[e.g.][]{1996JGR...10121639Z, 2001ApJ...551..495M}, to observed
line profiles, \citet{2002ApJ...574..412W, 2005ApJ...628L.143W}
estimate the wind ram pressures and mass loss rates of cool stars in
the solar neighbourhood.
Owing to the requirement of accurate ISM properties, which are only
available for the solar vicinity, the sample of observed stars is
currently small and rather heterogeneous, comprising single and binary
stars of different spectral types and luminosity classes.
For G and K main-sequence stars with X-ray fluxes $F_\mathrm{X}\lesssim
8\cdot10^5\,{\rm erg\cdot s^{-1}\cdot cm^{-2}}$,
\citeauthor{2005ApJ...628L.143W} find a correlation between the coronal
activity and the deduced mass loss rate per stellar surface area,
$\dot{M}/R^2\propto F_\mathrm{X}^{1.34\pm0.18}$.
Since the more active stars of the sub-sample have mass loss rates
lower than predicted by this power-law relation,
\citeauthor{2005ApJ...628L.143W} speculate that this may be due to a
change of the stellar magnetic field topology.

The $\dot{M}-F_\mathrm{X}$-relationships of \citet{2002ApJ...574..412W,
2005ApJ...628L.143W} have been applied to the rotational evolution of
cool stars and to their wind interaction with extra-solar planets.
\citet{2002ApJ...574..412W} extrapolate the mass loss history of
solar-like stars backward in time and suggest that the mass loss rate
of the young Sun may have been up to three orders of magnitude higher
than today.
Considering the `faint young Sun' paradox, that is the apparently
unchanged planetary temperatures despite an about $25\%$ less luminous
Sun about $3.8\,{\rm Gyr}$ ago \citep[e.g.][and references
therein]{2000GeoRL..27..501G, 2003JGRA.108l.SSH3S}, they find that the
cumulated mass loss of about $0.03\,{\rm M_{\sun}}$ can not account for
the $\sim 10\%$ difference in solar mass, which has been suggested to
solve this problem in terms of a more luminous, higher-mass young Sun.
\citet{2004A&A...425..753G} investigate the particle and thermal losses
from the upper atmospheres of `Hot Jupiters' and find that high stellar
mass loss rates have a significant influence on the evolution of the
planetary mass and radius.
The strength of stellar winds also affects the detectability of
extra-solar giant planets in radio wavelengths, since the emission
scales with the kinetic and magnetic energy flux on the planetary
magnetosphere \citep[e.g.][]{2005MNRAS.356.1053S,
2005A&A...437..717G}.
Owing to the high predicted stellar wind ram pressures, the
magnetospheric radio emission levels of close-by planets are expected
to be sufficiently high to become detectable with future instruments.

\citet{2002ApJ...574..412W, 2005ApJ...628L.143W} use scaled versions of
the solar wind to match observed and simulated line profiles.
With their principal quantity being the wind ram pressure, they presume
the terminal wind velocity of all stars to be solar-like, so that the
wind density is the only free fitting parameter.
The presumption of thermally driven winds with a unique terminal wind
velocity neglects, however, the influence of the magnetic field on the
acceleration and structuring of the outflow.
Coronal activity signatures and the magnetic braking of cool
main-sequence stars are indicative for surface magnetic flux, which is
generated by dynamo processes within the outer convection zone.
The magnetic field gives rise to the formation of hot X-ray emitting
coronal loops as well as to the acceleration of plasma escaping along
open field lines, and links ab initio the two quantities in the power
law suggested by \citet{2005ApJ...628L.143W}.

We consider the impact of magnetic fields in more detail by determining
wind ram pressures and mass loss rates in the framework of a magnetised
wind model.
Using the mass loss rates inferred by \citet{2002ApJ...574..412W,
2005ApJ...628L.143W} as empirical constraints for possible wind
scenarios, our working hypothesis is that the magnetic and thermal wind
properties of cool stars primarily depend on the stellar rotation
rate.
In Sect.\ \ref{moco}, we describe the magnetised wind model, the
principal quantities of the investigation, and the observational
constraints of the model parameters.
Section \ref{resu} comprises the analyses of characteristic wind
properties resulting from different wind scenarios, and the comparisons
of theoretical and empirical wind ram pressures and mass loss rates.
In Sect.\ \ref{disc}, we discuss our results and their consequences for
the evolution of stellar rotation and mass loss and for the
detectability of extra-solar planetary magnetospheres.
Our conclusions are summarised in Sect.\ \ref{conc}.

\section{Model considerations}
\label{moco}

\subsection{Magnetised wind model}
\label{mawimo}
We consider main-sequence stars with masses $0.2\,{\rm M_\odot}\le M\le
1.2\,{\rm M_\odot}$, radii $R\propto M^{0.8}$, and rotation rates
$0.6\,{\rm \Omega_\odot}\le \Omega\le 11.3\,{\rm \Omega_\odot}$; with
the solar rotation rate $\Omega_\odot= 2.8\cdot10^{-6}\,{\rm s^{-1}}$
the stellar rotation periods are between $2.3\,{\rm d}$ and $43\,{\rm
d}$.
The winds are determined in the framework of the magnetised wind model
of \citet{1967ApJ...148..217W}.
The properties of stationary, axisymmetric, and polytropic outflows are
specified through the wind temperature, $T_0$, the wind density,
$\rho_0$, and the radial magnetic field strength, $B_0$, at a reference
level, $r_0= 1.1\,{\rm R}$, close to the stellar surface.
The model solutions provide the flow velocity, $v_\mathrm{r,A}$, and
the density, $\rho_\mathrm{A}$, of the outflowing plasma at the
Alfv\'enic radius, $r_\mathrm{A}$, where the flow velocity equals the
local Alfv\'en velocity \citep{2005A&A...440..411H}.
The location of and the conditions at the Alfv\'enic point specify the
wind structure along a magnetic field line as well as the mass loss
rate \citep[see, e.g.,][]{1999isw..book.....L, 1999stma.book.....M}.

We assume that the entire stellar surface contributes to the wind.
Closed loop-like magnetic fields structures reduce the effective
surface area from which outflows can emanate and also influence the
flow structure in adjacent wind zones
\citep[e.g.][]{1968MNRAS.138..359M, 1974SoPh...34..231P,
1988ApJ...333..236K}.
The different magnetic field topologies encountered during a solar
activity cycles entail, for example, mass loss variations within a
factor of 2 \citep{1998csss...10..131W}.
Closed-field regions are however limited to a few stellar/solar radii
above the surface, well below the Alfv\'enic surface
\citep[e.g.][]{2002MNRAS.333..339J, 2002MNRAS.336.1364J}.
We expect that small-scale spatial and short-term temporal
inhomogeneities are averaged out with increasing distance from the star
and thus irrelevant for the sustainment of astrospheres.

\subsection{`Characteristic' wind ram pressure}
\label{cwrp}
\begin{figure}
\includegraphics[width=\hsize]{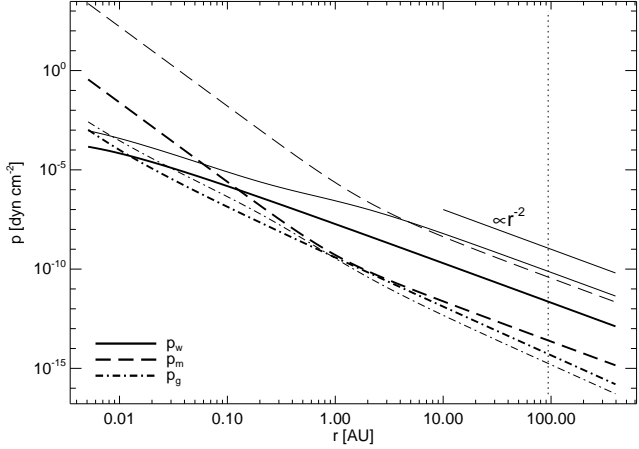}
\caption{Radial profiles of the ram pressure, $p_\mathrm{w}$, the
magnetic pressure, $p_\mathrm{m}$, and the thermal gas pressure,
$p_\mathrm{g}$, of magnetised winds in the equatorial plane.
For solar wind parameters (\emph{thick lines}), $T_0=
2.93\cdot10^6\,{\rm K}, n_0= 2.76\cdot10^6\,{\rm cm^{-3}}$, and $B_0=
3\,{\rm G}$, the magnetic and thermal pressures at the heliospheric
termination shock (\emph{dotted line}) are over two orders of magnitude
smaller than the wind ram pressure.
\emph{Thin lines} show the pressure profiles of a fast magnetic rotator
with a rotation period of six days and wind parameters $T_0=
3.4\cdot10^6\,{\rm K}, n_0= 6.67\cdot10^6\,{\rm cm^{-3}}, B_0=
243\,{\rm G}$.
At large distances from the star the wind ram pressure decreases
$\propto r^{-2}$.
Note that it is not the magnetic pressure which gives rise to the
magneto-centrifugal acceleration of the wind, but the stellar rotation
in conjunction with the plasma outflow along bent magnetic field
lines.}
\label{pprofiles.fig}
\end{figure}

The mass loss rate along a slender open magnetic flux tube is
\begin{equation}
d \dot{M} = F_\mathrm{m} d\sigma
\ ,
\label{defmalr}
\end{equation}
where $d\sigma= \sin \theta\, d\theta\, d\phi$ is the solid angle
occupied by the flux tube.
The mass flux per solid angle, $F_\mathrm{m}= \rho v_r r^2=
\rho_\mathrm{A} v_{r,{\rm A}} r_\mathrm{A}^2$, is constant along an
individual tube.  The rate of the momentum transport associated with
the mass flux,
\begin{equation}
d \dot{M} v_r = \rho v_r^2 r^2 d\sigma = p_\mathrm{w} d S
\ ,
\label{defmolr}
\end{equation}
is equivalent to the force which results from the wind ram pressure,
$p_\mathrm{w}= \rho v_r^2$, exerted on the local tube cross section, $d
S= r^2 d\sigma$.
At large distances from the star the outflow velocity converges to the
constant terminal velocity, $v_\infty$.
In contrast to the wind ram pressure, which decreases $\propto r^{-2}$
(Fig.\ref{pprofiles.fig}), the ram force per solid angle,
\begin{equation}
\frac{d\mathcal{F}_\mathrm{w}}{d\sigma}= p_\mathrm{w} r^2= F_\mathrm{m}
v_\infty= \mathrm{const.}
\ ,
\label{defrafo}
\end{equation}
is independent of the radius.
This quantity comprises both the wind density and (terminal) wind
velocity, and enables a characterisation of the mass loss rate of
individual stars.
In the case of spherically symmetric outflows, for which $d\sigma=
4\pi$, the ram force is $\mathcal{F}_\mathrm{w}= 4 \pi F_\mathrm{m}
v_\infty= \dot{M} v_\infty$.
To compare wind properties of stars with different stellar radii, we 
use the ram force per stellar surface area,
\begin{equation}
\mathcal{P}
=
\frac{\mathcal{F}_\mathrm{w}}{4\pi R^2}
= 
\frac{\dot{M} v_\infty}{4\pi R^2}
=
\frac{F_\mathrm{m} v_\infty}{R^2}
\ ,
\label{defcwrp}
\end{equation}
as the principal quantity of our investigation and shall refer to it as
the \emph{characteristic} wind ram pressure (CWRP).

\subsection{Empirical wind ram pressures}
\label{obco}
In their analysis of astrospherical absorption features,
\citet{2002ApJ...574..412W, 2005ApJ...628L.143W} adopt for all stars in
their sample solar-like winds with a unique terminal velocity of
$400\,{\rm km\cdot s^{-1}}$.
This assumption implies that the relative CWRP (in solar units),
\begin{equation}
\frac{\mathcal{P}}{\mathcal{P}_\odot}
=
\frac{\dot{M}}{\dot{M}_\odot}
\frac{v_\infty}{v_{\infty,\odot}}
\left( \frac{R}{R_\odot} \right)^{-2}
\ ,
\label{defrelcwrp}
\end{equation}
is equivalent to the stellar mass loss rate per surface area:
\begin{equation}
\left( \frac{\mathcal{P}}{\mathcal{P}_\odot} \right)_\mathrm{W}
=
\left( \frac{\dot{M}}{\dot{M}_\odot} \right)_\mathrm{W}
\left( \frac{R}{R_\odot} \right)^{-2}
\quad \textrm{for $v_\infty= v_{\infty,\odot}$}
\ .
\label{cwrpwood}
\end{equation}
In conjunction with individual stellar radii, the empirical mass loss
rates, $\dot{M}_\mathrm{W}$, given by \citet{2002ApJ...574..412W,
2005ApJ...628L.143W}, provide the observational constraints for our
theoretical CWRPs.
\begin{table}
\caption{Properties of cool main-sequence stars with observed mass loss
rates, taken from \citet{2005ApJ...628L.143W, 2005ApJS..159..118W};
$^a$ from \citet{1996ApJ...466..384D}; $^b$ from
\citet{1984ApJ...279..763N}.}
\begin{tabular}{cccccc}
\hline
star & sp.\,type & $P\,{\rm [d]}$ & $\dot{M}\,{\rm [\dot{M}_\odot]}$ & 
$\log L_X$ & $A\,{\rm [A_\odot]}$ \\
\hline
\object{Prox Cen} & M5.5 & 41.6 & $<0.2$ & 27.22 & 0.023 \\
\object{$\alpha$ Cen} & G2 / K0 & 29 / 43 & 2 & 27.70 & 2.22 \\
\object{$\epsilon$ Eri} & K2 & 11.7 & 30 & 28.32 & 0.61 \\
\object{61 Cyg A} & K5 & 35.4 & 0.5 & 27.45 & 0.46 \\
\object{$\epsilon$ Ind} & K4.5 & 22 & 0.5 & 27.39 & 0.56 \\
\object{36 Oph} & K1 / K1 & 20.7 / 22.9 & 15 & 28.34 & 0.88 \\
\object{EV Lac} & M3.5 & 4.38 & 1 & 28.99 & 0.123 \\
\object{70 Oph} & K0 / K5 & 19.7 / 22.9 & 100 & 28.49 & 1.32 \\
\object{$\xi$ Boo} & G8 / K4 & 6.31 / 11.94$^a$ & 5 & 28.90 & 1.0 \\
\object{61 Vir} & G5 & 32.7$^b$ & 0.3 & 26.87 & 1.0 \\
\hline
\object{Sun} & G2 & 26 & 1 & 27.30 & 1
\end{tabular}
\label{targets}
\end{table}

\subsection{Rotation-dependent wind parameters}
\label{roar}
The activity levels of cool stars increase with the stellar rotation
rate \citep[e.g.][]{1984ApJ...279..763N}, comprising enhanced thermal
and magnetic field values inside the coronae.
We take this aspect into account by assuming that the wind parameters
follow power-law relations, which are based on the solar reference
case.

The rates of increase of the thermal wind parameters\footnote{For
consistency reasons, we use the particle density, $n_0= \rho_0
N_\mathrm{A} / \mu$, to specify the wind condition at the reference
level, with the mean molecular weight $\mu$ and the Avogadro number
$N_\mathrm{A}$.}
\begin{equation}
T_0= T_{0,\odot} \left( \frac{\Omega}{\Omega_\odot} \right)^{n_T}
\quad \textrm{and} \quad
n_0= n_{0,\odot} \left( \frac{\Omega}{\Omega_\odot} \right)^{n_n}
\label{trhopowlaws}
\end{equation}
at the reference level are specified through the power-law indices
$n_T$ and $n_n$, respectively.
Polarimetric observations indicate that the magnetic flux depends on 
the stellar rotation rate \citep[e.g.][]{1991LNP...380..389S}.
Taking different stellar radii (here, $R\propto M^{0.8}$) into account,
the magnetic field strengths are taken to follow the power-law relation
\begin{equation}
B_0= B_{0,\odot} \left( \frac{M}{M_\odot} \right)^{-1.6} 
\left( \frac{\Omega}{\Omega_\odot} \right)^{n_\Phi}
\ ,
\label{bpowlaws}
\end{equation}
which implies that for a given stellar rotation rate lower-mass stars 
have higher average field strengths.

The assumption of polytropic magnetised winds whose wind parameters
follow simple power-laws may be simplistic, but in view of the
currently limited observational constraints a more sophisticated model
appears inappropriate.

\subsubsection{Observational constraints}
Since the thermal wind properties of cool stars cannot be observed
directly, we follow the hypothetical assumption that closed coronal
magnetic field regions may serve as proxies to constrain the increase
of the temperature and density with the stellar rotation rate.
Analysing the dependence of stellar X-ray luminosities,
\citet{2003ApJ...599..516I} infer a density power-law index $n_n\approx
0.6$, which implies for stars rotating ten times faster than the Sun
(i.e.\ close to the X-ray saturation limit) about four times higher
coronal densities.
Differential emission measures of rapidly rotating stars indicate a
large fraction of plasma with temperatures of $\sim 10^7\,{\rm K}$, in
addition to solar-like coronal plasma components with temperatures
$\gtrsim 10^6\,{\rm K}$ \citep[e.g.][and references therein]{
2003SSRv..108..577F}.
A coronal temperature of $10\,{\rm MK}$ for a rapidly rotating star
like AB Dor ($\Omega\simeq 50\,{\rm \Omega_\odot}$) would imply a
temperature power-law index of $n_T\simeq 0.5$.
Following the solar paradigm, it is more likely that the temperature of
stationary stellar winds are characterised by the low-temperature
plasma component, implying values $0\lesssim n_T< 0.5$.

The increase of the magnetic flux with the stellar rotation rate is
constrained through direct magnetic flux measurements \citep[e.g.][and
references therein]{1991LNP...380..389S, 2001ASPC..223..292S} and,
indirectly, through empirical activity-rotation-age relations.
Due to the braking effect of magnetised winds, solar-like single stars
spin down during the course of their main-sequence evolution.
In conjunction with the \citet{1967ApJ...148..217W}-wind model,
empirical spin-down timescales \citep[e.g.][]{1972ApJ...171..565S}
imply a linear relationship between the stellar magnetic flux and the
rotation rate, but sub-linear values cannot be ruled out either, owing
to the impact of non-uniform surface magnetic field distributions
\citep{2005A&A...444..661H}.
In contrast, \citet{2001ASPC..223..292S} suggests a super-linear
dependence, with a power-law index of $n_\Phi= 1.2$, based on
polarimetric observations of K and M dwarfs.
Depending on the underlying stellar sample, higher values may be
possible.
Excluding rapid rotators in the saturated regime,
\citet{2003ApJ...590..493S}, for example, suggest the value $n_\Phi=
2.8\pm0.3$.

\subsection{Solar reference case}
\label{src}
For our model to reproduce solar wind conditions observed at earth
orbit ($v_r\simeq 400\,{\rm km\cdot s^{-1}}, n\footnote{The average
proton density in the slow solar wind is about $7\,{\rm cm^{-3}}$.
Owing to the charge-neutrality of the plasma, we double this value to
account for the free electrons. This is consistent with the mean
molecular weight $\mu= 0.5$} \simeq 14\,{\rm cm^{-3}}, T\simeq
2\cdot10^5\,{\rm K}, B\simeq 5\cdot10^{-5}\,{\rm G}$), the coronal
values have to be $T_{0,\odot}= 2.93\cdot10^6\,{\rm K}, n_{0,\odot}=
2.76\cdot10^6\,{\rm cm^{-3}}, B_{0,\odot}= 3\,{\rm G}$, the polytropic
index $\Gamma= 1.22$, and the mean molecular weight $\mu= 0.5$
\citep[cf.][]{1967ApJ...148..217W, 1985A&A...152..121S}.
The resulting mass flux is $F_\mathrm{m,\odot}= 1.04\cdot10^{11}\,{\rm
g\cdot s^{-1}\cdot sr^{-1}}$ and the solar mass loss rate
$\dot{M}_\odot= 4 \pi F_\mathrm{m,\odot}= 1.31\cdot10^{12}\,{\rm g\cdot
s^{-1}}= 2.07\cdot10^{-14}\,{\rm M_\odot\cdot yr^{-1}}$.
With increasing distance from the Sun, the wind velocity converges to
the value $v_{\infty,\odot}= 443\,{\rm km\cdot s^{-1}}$, which yields
the constant wind ram force per solid angle
$\mathcal{F}_\mathrm{w,\odot}/(4\pi)= 4.60\cdot10^{18}\,{\rm dyn\cdot
sr^{-1}}$.
With the solar radius $R_\odot= 6.96\cdot10^{10}\,{\rm cm}$, the CWRP
of the Sun is $\mathcal{P}_\odot= 9.51\cdot10^{-4}\,{\rm dyn\cdot
cm^{-2}}$.

The interaction with the local interstellar medium confines the
expansion of the solar wind within a system of shock fronts, wherein
velocities are braked down to sub-sonic values.
At the distance of the heliospheric termination shock ($\approx
94\,{\rm AU}$, \citealt{2005Sci...309.2017S}) the calculated solar wind
has virtually obtained the terminal velocity ($437\,{\rm km\cdot
s^{-1}}= 0.987\cdot v_{\infty,\odot}$) and exerts a ram pressure
($p_\mathrm{w,TS} \simeq 2\cdot10^{-12}\,{\rm dyn\cdot cm^{-2}}$, cf.\
Fig.\ \ref{pprofiles.fig}), which is in agreement with the canonical
value of the pressure of the local interstellar medium
\citep[e.g.][]{1986AdSpR...6...27A}.

\section{Results}
\label{resu}

\subsection{Comparison of different wind scenarios}
\label{comp}
\begin{figure}
\includegraphics[width=\hsize]{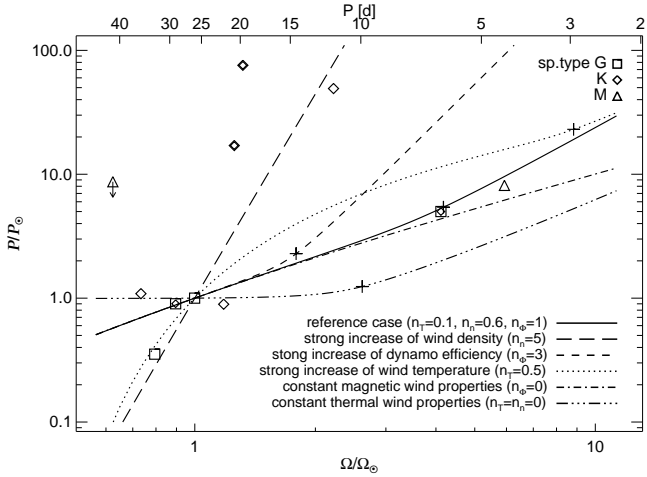}
\caption{Characteristic wind ram pressure (in solar units) of $1\,{\rm
M_{\sun}}$ main-sequence stars subject to different wind scenarios
(i.e.\ sets of power-law indices $n_T, n_n$, and $n_\Phi$).
\emph{Crosses} along the curves mark the transition between the regimes
of slow and fast magnetic rotators, and \emph{symbols} show the values
inferred by \protect \citet{2005ApJ...628L.143W} for main-sequence
(single and binary) stars of different spectral types.
}
\label{mlmomwd.fig}
\end{figure}

An increase of thermal and/or magnetic wind parameters with the stellar
rotation rate increases the CWRPs (Fig.\ \ref{mlmomwd.fig}).
The character of the increase depends on whether the star is a
\emph{slow} or a \emph{fast magnetic rotator}\footnote{We follow the
terminology of \citet{1976ApJ...210..498B} and consider stars to be
fast magnetic rotators, if the Michel velocity, which quantifies the
impact of magnetic fields on the wind acceleration
\citep{1969ApJ...158..727M}, is larger than the terminal wind velocity
determined in the absence of magneto-rotational effects
\citep[cf.][]{1980ApJ...242..723N, 1999isw..book.....L}.}.
The winds of slow magnetic rotators are driven by thermal pressure
gradients and gain energy from the enthalpy of the hot plasma.
In the regime of fast magnetic rotators, stellar winds are
predominantly accelerated by magneto-centrifugal driving.
The outflowing plasma tries to conserve its angular momentum but is
forced into faster rotation by the tension force of the bent magnetic
field lines.
This slingshot effect gives rise to a Poynting flux, which transfers
energy from the stellar rotation into the wind.

\begin{figure}
\includegraphics[width=\hsize]{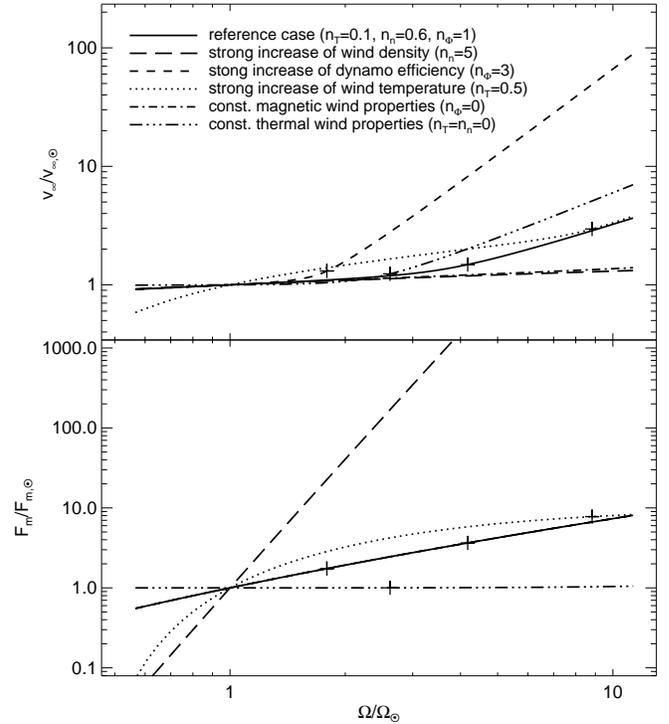}
\caption{Terminal velocities (\emph{top}) and mass fluxes
(\emph{bottom}) of winds of $1\,{\rm M_{\sun}}$ main-sequence stars
subject to different wind scenarios.
In the cases of constant ($n_\Phi= 0$, \emph{dashed dotted}) and
strongly increasing ($n_\Phi= 3$, \emph{short dashed}) dynamo
efficiencies the mass fluxes are identical with those of the reference
case (\emph{solid}), since they are based on the same thermal wind
parameters.
However, the individual transitions between the slow and fast magnetic
rotator regimes (\emph{crosses}) take place at different rotation
rates.
}
\label{fmtv.fig}
\end{figure}

In the regime of slow magnetic rotators, the terminal wind velocities
are comparable to the solar-like surface escape velocities of
main-sequence stars (Fig.\ \ref{fmtv.fig}).
Except in the case of high wind temperatures, the CWRPs only depend on
the mass flux, which scales almost linearly with the wind density.
Higher wind temperatures entail higher mass fluxes and CWRPs, yet the
increase is limited by the basic requirement of subsonic coronal flow
velocities.
The terminal wind velocities of fast magnetic rotators are larger than
the solar value.
The mass flux, however, is still determined by the thermal wind
parameters.
Owing to its intrinsic dependence on the stellar rotation, the dominant
magneto-centrifugal driving causes an increase of the terminal wind
velocity and CWRP with the stellar rotation rate even if the wind
parameters are constant.
The transition between the slow and the fast magnetic rotator regimes
depends on the relative contribution of thermal and magneto-centrifugal
driving to the overall acceleration of the wind.
The higher the dynamo efficiency, $n_\Phi$, the stronger the
magneto-centrifugal driving, and the lower the rotation rate of the
transition between the two regimes.
Higher wind temperatures increase the thermal driving and shift the
transition to higher rotation rates.

We have compared self-consistently determined CWRPs of different wind
scenarios with the observational constraints (Table \ref{targets}).
At present, the data set for main-sequence stars consists of three
groups: five slowly rotating targets with solar-like CWRPs
(\object{Sun}, \object{$\alpha$ Cen}, \object{$\epsilon$ Ind},
\object{61 Cyg A}, \object{61 Vir}), three moderately rotating targets
of spectral type K with CWRPs more than ten times the solar value
(\object{$\epsilon$ Eri}, \object{36 Oph}, \object{70 Oph}), and two
rapidly rotating targets (\object{$\xi$ Boo}, \object{EV Lac}) with
5-10 times larger CWRPs than the Sun; the value for \object{Proxima
Cen} is an upper limit.
We first focus on the slowly and rapidly rotating targets and consider
the group of moderately rotating K dwarfs in Sect.\ \ref{kdwarfs}.

Assuming a linear dynamo efficiency, $n_\Phi= 1$, different thermal
wind scenarios are (within the given observational error margins)
consistent with the constraints set by the slowly and rapidly rotating
stars (Fig.\ \ref{mlmomwd.fig}).
Rotation-independent thermal wind properties appear less likely, since
the resulting constant CWRPs at small rotation rates are inconsistent
with the general trend of CWRPs increasing with stellar rotation.
Good agreement results from moderately increasing thermal wind
parameters, here $n_T= 0.1$ and $n_n= 0.6$ (reference case), but the
empirical constraints do not conclusively exclude the scenario either,
in which the wind temperature follows the high-temperature coronal
plasma component (i.e.\ $n_T= 0.5$).
In the latter scenario, the strong increase of CWRPs at low rotation
rates, which then becomes flatter for higher rotation rates, appears
promising to account for the high values of the group of K dwarfs.  But
we find that even for $n_T\gg 0.5$ the CWRPs are not sufficiently high.
The only scenario consistent with these targets is based on a very
strong increase of the coronal density with the stellar rotation rate.
For stars close to the X-ray saturation limit, rotating about ten times
faster than the Sun, values of $n_n\gtrsim 5$ imply very high wind
densities and CWRPs which are inconsistent with the lower values
inferred for the rapidly rotating targets.
The scenario of wind ram pressures scaling exclusively with the wind
density corresponds to the approach of \citet{2002ApJ...574..412W}.

Retaining the thermal power-law indices $n_T= 0.1$ and $n_n= 0.6$, the
theoretical CWRPs are consistent with the constraints set by the slowly
and rapidly rotating targets for dynamo efficiencies $0< n_\Phi\lesssim
1.5$.
For $n_\Phi< 1$ all stars considered here are slow magnetic rotators,
whereas for super-linear dynamo efficiencies the two rapidly rotating
targets, \object{$\xi$ Boo} and \object{EV Lac}, are in the fast
magnetic rotator regime.
A rather high but previously suggested dynamo efficiency of $n_\Phi\sim
3$ locates the transition between slow and fast magnetic rotators at
rotation rates similar to those of the K dwarf group.
Yet due to their moderate rotation, the magneto-centrifugal driving is
too inefficient and cannot produce CWRPs sufficiently high to match the
empirical values.

None of the scenarios above is capable to account for all of the
empirical constraints, and all but one scenario are incapable to match
the high CWRPs of the group of moderately rotating K dwarfs.
With highest values occurring at intermediate rotation rates, it is
unlikely that the wind ram pressures of the three groups can be
described using simple power-law relations for the thermal and magnetic
wind properties.
The group of moderately rotating K dwarfs shows high CWRP, which would
require a sudden change of wind properties with the stellar rotation
rate.
But assuming a continuous dependency makes it difficult to yield lower
CWRPs at higher stellar rotation rates, as suggested by \object{$\xi$
Boo} and \object{EV Lac}.
In contrast, a weak dependence on the stellar rotation rate enables to
match the values of the rapidly rotating targets, but cannot account
for the high CWRPs of the K dwarf group.
This suggests that either the group of rapidly rotating targets or the
group of moderately rotating K dwarfs are peculiar in terms of stellar
wind properties.

\subsection{The group of moderately rotating K dwarfs}
\label{kdwarfs}
In the framework of our wind model with rotation-dependent thermal and
magnetic wind parameters, the CWRPs of the three moderately rotating
targets \object{36 Oph} ($\mathcal{P}/\mathcal{P}_\odot= 17$),
\object{70 Oph} ($\mathcal{P}/\mathcal{P}_\odot= 75$), and
\object{$\epsilon$ Eri} ($\mathcal{P}/\mathcal{P}_\odot= 49$), could
only be accounted for by a strong increase of wind densities with
stellar rotation rate.
For otherwise solar (i.e.\ reference) wind parameters, the wind
densities required to produce the empirical CWRPs are about 25 times
(36 Oph), 82 times (70 Oph), and 65 times the solar value.
Yet, according to analyses of recent X-ray observations
\citep{2006ApJ...643..444W}, the coronae of the three targets are
solar-like as far as densities are concerned.
We investigate what other wind conditions may cause high CWRPs, and
whether these are consistent with observations.

\begin{figure}
\includegraphics[width=\hsize]{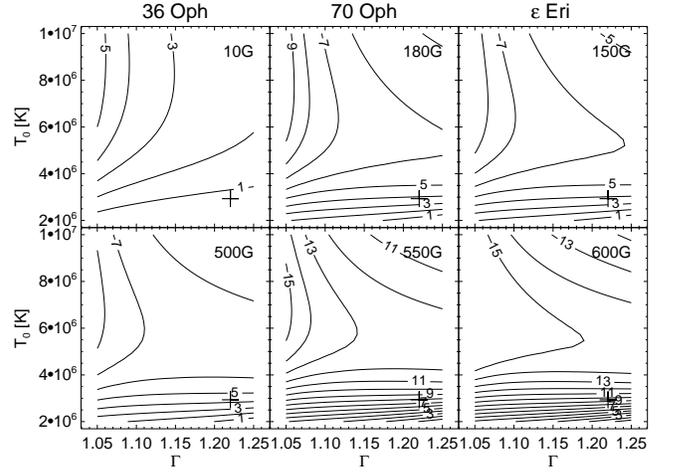}
\caption{Theoretical characteristic wind ram pressures (in solar units)
of the moderately rotating K dwarf targets as function of the wind
temperature, $T_0$, at the coronal base and the polytropic index,
$\Gamma$; the \emph{cross} marks solar reference values.
The magnetic field strengths, $B_0$ (\emph{labels}), are lower and
upper limits of observed field strength ranges.
The stellar model parameters are for \object{36 Oph}: $M= 0.63\,{\rm
M_{\sun}}, r_0= 0.76\,{\rm R_{\sun}}, \Omega= 3.5\cdot10^{-6}\,{\rm
s^{-1}}$; \object{70 Oph}: $M= 0.8\,{\rm M_{\sun}}, r_0= 0.94\,{\rm
R_{\sun}}, \Omega= 3.7\cdot10^{-6}\,{\rm s^{-1}}$; \object{$\epsilon$
Eri}: $M= 0.73\,{\rm M_{\sun}}, r_0= 0.86\,{\rm R_{\sun}}, \Omega=
6.2\cdot 10^{-6}\,{\rm s^{-1}}$.
The wind density at the reference level is $\rho_0= 2.76\cdot10^6\,{\rm
cm^{-3}}$ and the mean molecular weight $\mu= 0.5$.}
\label{kdwarfs.fig}
\end{figure}

Regarding their rather slow rotation, the three targets have
unexpectedly high surface-averaged magnetic field strengths: $ (0) -
500\,{\rm G}$ for \object{36 Oph} \citep{1984ApJ...276..286M,
1997MNRAS.284..803S}, $180-550\,{\rm G}$ for \object{70 Oph}
\citep{1980ApJ...236L.155R, 1984ApJ...276..286M}, and $150-600\,{\rm
G}$ for \object{$\epsilon$ Eri} \citep{1984ApJ...276..286M,
1995ApJ...439..939V, 1997A&A...318..429R}.
The lower values are considered to be more reliable, but owing to
magnetic filling factors smaller one peak field strengths may locally
reach $1-3\,{\rm kG}$.
We determine CWRPs for the three targets as functions of both the wind
temperature at the reference level and the polytropic index.
The wind density at the reference level is taken to be $\rho_0=
2.76\cdot10^6\,{\rm cm^{-3}}$, whereas individual stellar rotation
rates, radii, and masses are taken from Table \ref{targets},
\citet{2005ApJS..159..118W}, and the approximation $M\propto R^{5/4}$
(cf. Sect.\ \ref{mawimo}), respectively.
The calculations are carried out for the lower and upper limits of each
magnetic field strength range given above (Fig.\ \ref{kdwarfs.fig}).
For drastic thermal wind conditions, like base temperatures
$T_0\lesssim 10^7\,{\rm K}$ and $\Gamma\gtrsim 1$ (implying almost
isothermal outflows), CWRPs increase to about 5-15 times the solar
value.
Despite the large observational uncertainties of a factor of two
\citep{2002ApJ...574..412W}, the resulting CWRPs are insufficient to
achieve conclusive agreements with the empirical values.

\begin{figure}
\includegraphics[width=\hsize]{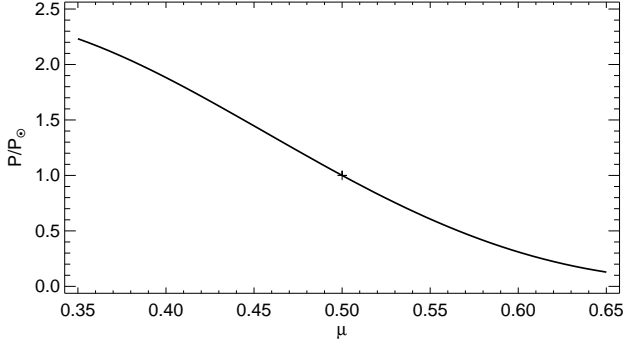}
\caption{Dependence of the characteristic wind ram pressure (in solar
units) on the mean molecular weight, assuming solar reference values
($M= 1\,{\rm M_{\sun}}, \Omega= 2.8\cdot10^{-6}\,{\rm s^{-1}}, r_0=
1.1\,{\rm R_{\sun}}, B_0= 3\,{\rm G}, T_0= 2.93\cdot10^6\,{\rm K}, n_0=
2.76\cdot10^6\,{\rm cm^{-3}}, \Gamma= 1.22$).}
\label{muscan.fig}
\end{figure}

The Chandra observations of \citet{2006ApJ...643..444W} show abundance
anomalies for some of the K dwarf targets in the form of a FIP effect,
that is the relative abundances (with respect to photospheric values)
of elements with low first ionisation potentials are enhanced compared
to high-FIP elements.
The element abundances determine the mean molecular weight of the
coronal plasma, and could thus have an influence on the wind
acceleration mechanisms and mass loss rate.
Assuming a fully ionised plasma, solar photospheric abundances
\citep{1998SSRv...85..161G} yield $\mu_\mathrm{\odot,ph}\approx 0.61$.
In the solar wind, elements with $\rm{FIP}< 10\,{\rm eV}$ are about 4.5
times overabundant and elements with $10\,{\rm eV}< \rm{FIP} <
11.5\,{\rm eV}$, in particular C and S, about two times overabundant
\citep{1998SSRv...85..241G, 1999SSRv...87...55R}; in the slow wind, He
($\rm{FIP}\sim 25\,{\rm eV}$) is about two times underabundant
\citep{1998SSRv...85..241G}.
These abundance differences result in $\mu_\mathrm{\odot,wind}\approx
0.57$.
\citet{2006ApJ...643..444W} determined the relative abundances of the
major constituents of the K dwarf coronae.
If we assume that in stellar winds He is generally underabundant, the
mean molecular weights are $\mu_\mathrm{36\,Oph (A/B)}\approx 0.53,
\mu_\mathrm{70\,Oph (A)}\approx 0.58, \mu_\mathrm{70\,Oph (B)}\approx
0.54$, and $\mu_\mathrm{\epsilon\,Eri}\approx 0.54$, that is $\lesssim
10\%$ smaller than the solar value.
However, the dependence of the CWRP on the mean molecular weight is
small (Fig.\ \ref{muscan.fig}), and this effect therefore rather
marginal.

In summary, high magnetic field strengths, high wind temperatures, high
heating rates and low mean molecular weights do, in principle, increase
the CWRPs of cool stars.
Yet, if the energy flux into open and closed magnetic field structures
is similar, then we expect that the extreme coronal conditions required
to yield CWRPs nearly two orders of magnitude larger than the Sun would
imprint distinctive signatures on stellar X-ray properties.
Since these signatures are not discernible in the case of the
moderately rotating K dwarfs, their high CWRPs are peculiar.

\subsection{Dependence on spectral type}
\begin{figure}
\includegraphics[width=\hsize]{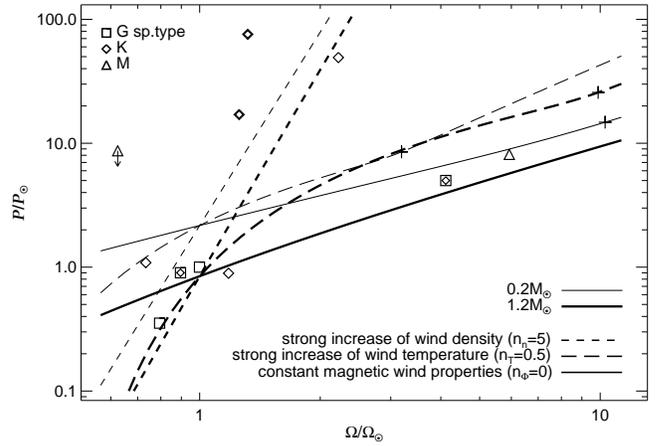}
\caption{Characteristic wind ram pressures of $1.2\,{\rm M_{\sun}}$
(\emph{thick lines}) and $0.2\,{\rm M_{\sun}}$ (\emph{thin lines})
main-sequence stars, subject to different wind scenarios.
\emph{Crosses} mark the transition between the slow and the fast
magnetic rotator regimes, \emph{symbols} show the empirical values
given by \protect \citet{2005ApJ...628L.143W}.}
\label{mlmomwd_stc.fig}
\end{figure}

For comparable wind parameters and rotation rates, the CWRPs of
lower-mass main-sequence stars are higher than for solar-mass stars.
Over the mass range $0.2-1.2\,{\rm M_{\sun}}$, the difference is
typically smaller than about half an order of magnitude (Fig.\
\ref{mlmomwd_stc.fig}).
This holds for both the slow and the fast magnetic rotator regime, so
it is unrelated to the higher magnetic field strengths for lower mass
stars, adapted to retain the total magnetic flux (cf.\ Sect.\
\ref{roar}).
The increase is caused by the scaling of $\mathcal{P}$ with the inverse
surface area (i.e.\ $\propto R^{-2}$).
The mass fluxes (per solid angle) of lower-mass stars are, in fact,
significantly smaller than the solar value.
In the case of the reference wind scenario, the CWRPs and mass loss
rates follow approximately broken power-laws (Table \ref{cwrpfit}),
whose parameters reflect the result above.
\begin{table*}
\caption{Power-law approximations for the CWRP, $\mathcal{P}=
\bar{\mathcal{P}} \left( \Omega / \Omega_\odot
\right)^{n_\mathcal{P}}$, the mass flux (per solid angle),
$F_\mathrm{m}= \dot{M}/(4\pi)= \bar{F}_\mathrm{m} \left( \Omega /
\Omega_\odot \right)^{n_F}$, and the terminal wind velocity, $v_\infty=
\bar{v}_\infty \Omega^{n_v}$, of main-sequence stars, assuming a wind
scenario with $n_T= 0.1, n_n= 0.6$, and $n_\Phi= 1$.
The transition between the slow and the fast magnetic rotator regime,
SMR and FMR, respectively, takes place at $\Omega_\mathrm{s/f}$.}
\begin{tabular}{cc|cccccc|cccccc}
\hline
$M$ & $\Omega_\mathrm{s/f}$ & 
\multicolumn{6}{c}{SMR ($\Omega< \Omega_\mathrm{s/f}$)} &
\multicolumn{6}{c}{FMR ($\Omega> \Omega_\mathrm{s/f}$)}
\\
$[M_{\sun}]$ & $[\Omega_\odot]$ &
$\bar{\mathcal{P}}\,{\rm [\mathcal{P}_\odot]}$ & $n_{\mathcal{P}}$ &
$\bar{F}_\mathrm{m}\,{\rm [F_\mathrm{m,\odot}]}$ & $n_F$ &
$\bar{v}_\infty\,{\rm [v_{\infty,\odot}]}$ & $n_v$ &
$\bar{\mathcal{P}}\,{\rm [\mathcal{P}_\odot]}$ & $n_{\mathcal{P}}$ &
$\bar{F}_\mathrm{m}\,{\rm [F_\mathrm{m,\odot}]}$ & $n_F$ &
$\bar{v}_\infty\,{\rm [v_{\infty,\odot}]}$ & $n_v$
\\
\hline
1.2 & 4.39 & 0.83 & 1.21 & 1.15 & 0.99 & 0.97 & 0.21 & 0.40 & 1.73 &
1.41 & 0.82 & 0.38 & 0.90 \\
1.0 & 4.18 & 0.99 & 1.14 & 0.99 & 0.94 & 1.00 & 0.20 & 0.45 & 1.73 &
1.18 & 0.79 & 0.38 & 0.93 \\
0.8 & 3.90 & 1.18 & 1.08 & 0.80 & 0.89 & 1.04 & 0.19 & 0.52 & 1.72 &
0.92 & 0.76 & 0.39 & 0.95 \\
0.6 & 3.54 & 1.43 & 1.01 & 0.58 & 0.84 & 1.09 & 0.17 & 0.61 & 1.72 &
0.66 & 0.73 & 0.41 & 0.99 \\
0.4 & 3.05 & 1.75 & 0.94 & 0.35 & 0.78 & 1.14 & 0.16 & 0.77 & 1.72 &
0.38 & 0.70 & 0.46 & 1.02 \\
0.2 & 2.30 & 2.17 & 0.85 & 0.13 & 0.71 & 1.23 & 0.14 & 1.10 & 1.72 &
0.14 & 0.65 & 0.59 & 1.07
\end{tabular}
\label{cwrpfit}
\end{table*}
For K dwarfs with masses $\sim 0.7\,{\rm M_{\sun}}$, the increase of
the CWRP is clearly insufficient to account for the high observed
values of the peculiar group of moderately rotating targets.

\subsection{Inclination effects}
The wind ram pressure of magnetised winds is intrinsically
latitude-dependent, since plasma emanating along open magnetic field
lines at high latitudes experience a weaker magneto-centrifugal driving
than outflows in the equatorial plane.
Inferred mass loss rates based on the assumption of spherical symmetric
outflows may misestimate the actual value if the inclination between
the line-of-sight and the rotation axis is not taken into account.
We analyse the impact of this effect by determining latitude-dependent
CWRPs.
The applied model is an extension of the
\citeauthor{1967ApJ...148..217W}-formalism to non-equatorial latitudes,
assuming that the poloidal magnetic field component is radial, so that
the spiralling field lines are located on conii with constant opening
angles, whose tips are located in the centre of the star
\citep{2005A&A...440..411H}.

\begin{figure}
\includegraphics[width=\hsize]{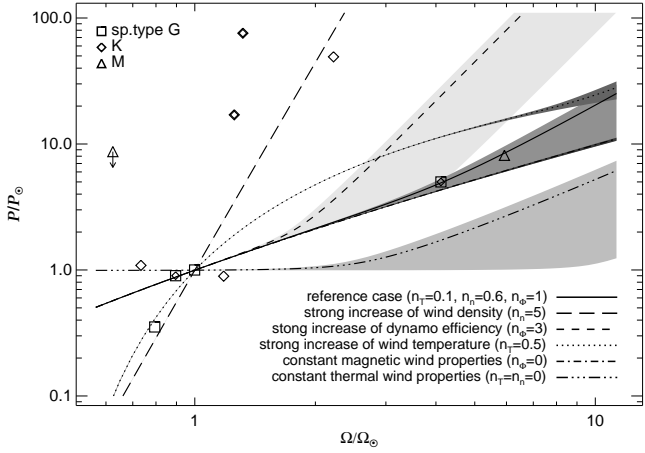}
\caption{Characteristic wind ram pressures of outflowing plasma with
different inclinations to the stellar rotation axis.
Based on uniform surface magnetic field distributions, higher (lower)
values within a \emph{shaded region} are associated with equatorial
(high) latitudes.
The \emph{lines} show surface-averaged mean values, that is the values
one would statistically expect if the inclination is unknown.
\emph{Symbols} indicate the values and spectral types of the empirical
data of \protect \citet{2005ApJ...628L.143W}.}
\label{mlmom.fig}
\end{figure}

Disregarding possible anisotropies caused by closed coronal magnetic
field structures, the thermally driven winds of slow magnetic rotators
are virtually spherical symmetric and inferred CWRPs independent of
inclination effects.
In contrast, in the regime of fast magnetic rotators, possible CWRPs
range over up to an order of magnitude, depending on the underlying
wind scenario (Fig.\ \ref{mlmom.fig}).
Small inclinations between the rotation axis and the line-of-sight
typically imply smaller values.
Considering the larger surface area at lower latitudes, the expectation
values are closer to the higher CWRPs determined in the equatorial
plane.

The present analysis is based on uniform magnetic field distributions.
Yet observations of rapidly rotating stars frequently show non-uniform
surface brightness and magnetic field distributions in the form of
spots concentrations at high latitudes \citep[e.g.][and references
therein]{2002AN....323..309S}, which alter the latitudinal variation of
the wind structure.
Magnetised winds of rapidly rotating stars furthermore show a
collimation of open magnetic field lines toward the rotation axis,
which causes an additional latitude-dependence of the wind structure
\citep[e.g.][]{2000A&A...356..989T}.
Such anisotropies increase the range of possible CWRPs and the
uncertainty of inferred values.

\subsection{Rotation-activity-relationship}
\begin{figure}
\includegraphics[width=\hsize]{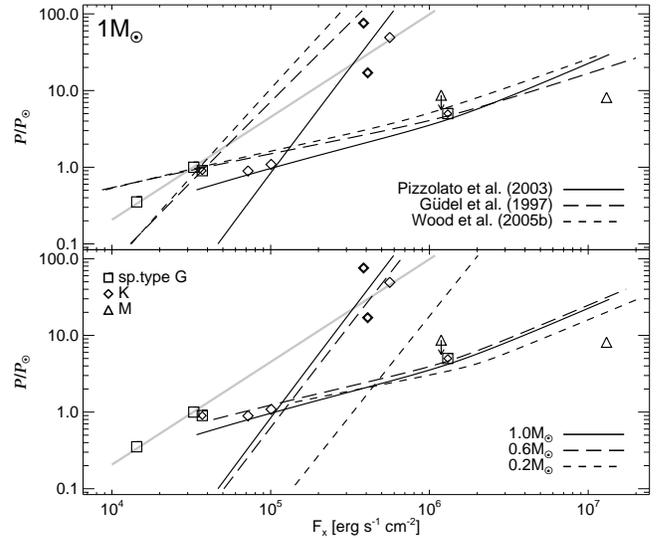}
\caption{Characteristic wind ram pressures (in solar units) as function
of the stellar X-ray flux, assuming the reference ($n_T= 0.1, n_n= 0.6,
n_\Phi= 1$) and the high wind density ($n_T= 0.1, n_n= 5, n_\Phi= 1$,
\emph{steep curves}) scenarios.
Different empirical rotation-activity-relations result in different
locations and slopes of the curves for a $1\,{\rm M_{\sun}}$ star
(\emph{top}).
Using the rotation-activity relationship based on data by
\protect\citet{2003A&A...397..147P}, the dependence on spectral type
(i.e.\ stellar mass) shifts the values of lower-mass stars to higher
X-ray fluxes (\emph{bottom}).
The \emph{gray} line marks the X-ray flux-mass loss-relation, $\dot{M}
\propto F_X^{1.34}$, suggested by
\protect\citet{2005ApJ...628L.143W}.}
\label{mlmfxwd.fig}
\end{figure}

The activity level of cool stars is typically quantified through their
coronal X-ray emission; \citet{2002ApJ...574..412W,
2005ApJ...628L.143W} correlate the inferred mass loss rates with
observed X-ray fluxes.
For cool stars with rotation periods longer than about two days, X-ray 
fluxes increase with the stellar rotation rate, following (on average) 
a power-law relation, 
\begin{equation}
F_X= \bar{F}_X \Omega^{n_X}
\ .
\label{deffxom}
\end{equation}
The empirical values $\bar{F}$ and $n_X$ depend on the spectral type 
and size of the underlying stellar sample.
We demonstrate the impact of these quantities on the mapping of our
theoretical CWRPs onto the $F_X$-$\mathcal{P}$-plane by comparing
several relationships (Table \ref{fxnx.tbl}).
\begin{table}
\caption{Coefficients and power-law indices of empirical
rotation-activity-relationships.
The mass dependence has been determined by fitting power laws to the
characteristic values of each mass bin in the data set given by
\citet{2003A&A...397..147P}.}
\label{fxnx.tbl}
\begin{tabular}{ccc}
\hline
$\lg \bar{F}_X\,{\rm [erg\cdot s^{-1}\cdot cm^{-2}]}$ & $n_x$ & sample \\
\hline
$15.51 + 0.623 \left( \frac{M}{M_{\sun}} \right) - 2 \lg \left(
\frac{R}{R_{\sun}} \right)$ & $2$ & 259 F\,G\,K\,M stars$^a$ 
\\
$19.25 - 2 \lg \left( \frac{R}{R_{\sun}} \right)$ & $2.64$ & 9 G stars$^b$
\\ 
$17.88$ & $2.4$ & 19 F\,G stars$^c$
\end{tabular}
\newline
{\small $^a$ \citet{2003A&A...397..147P}; $^b$
\citet{1997ApJ...483..947G}; $^c$ \citet{2005ApJS..159..118W} }
\label{rotactrels}
\end{table}
Larger $\bar{F}_X$-values shift the curves of theoretical CWRPs to
higher coronal X-ray fluxes, whereas smaller power-law indices $n_X$
cause steeper curves, since $d \ln \mathcal{P} / d \ln F_X= (d \ln
\mathcal{P} / d \ln \Omega) / n_X$ (Fig.\ \ref{mlmfxwd.fig}).
The latter effect would ease the need for a strong increase of the wind
density with the stellar rotation.
Depending on the applied rotation-activity relation, values $n_n\simeq
2.2-3.1$ yield CWRPs which are in agreement with the moderately
rotating K dwarfs and the power-law relationship suggested by
\citet{2005ApJ...628L.143W}.
Yet the offsets caused by different $\bar{F}_X$ result in
inconsistencies between theoretical and empirical CWRPs.

The range of possible locations and slopes of CWRP curves resulting
from different empirical rotation-activity relations make it difficult
to associate inferred mass loss rates with a systemic change of stellar
wind parameters.
The rotation-activity relationships are statistical relations based on
a (relative) large number of stars.
For individual objects,  deviations from the mean value can be
significant and represent an additional source of uncertainty in the
analysis of stellar wind ram pressures.
Furthermore, X-ray fluxes of cool stars are typically not constant, but
may vary in the course of an activity cycle, in the case of the Sun by
an order of magnitude.

\subsection{Stellar mass loss rates}
\begin{figure}
\includegraphics[width=\hsize]{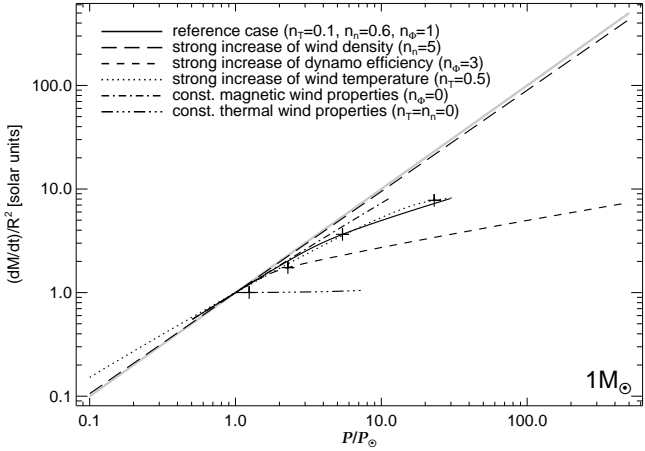}
\caption{Self-consistent mass loss rates per surface area,
$\dot{M}/R^2$ (in solar units), as a function of the relative
characteristic wind ram pressure, $\mathcal{P} / \mathcal{P}_\odot$.
Subject to the condition $v_\infty= v_{\infty,\odot}$, the latter
quantity is equivalent to the empirical mass loss rates per surface
area given by \protect \citeauthor{2002ApJ...574..412W}.
\emph{Crosses} mark the transition between the regimes of slow and fast
magnetic rotators; the \emph{gray line} indicates identity between
theoretical and observed mass loss rates.}
\label{maslos.fig}
\end{figure}

Following Eqs. (\ref{defrelcwrp}) and (\ref{cwrpwood}), a match between
theoretical and empirical CWRPs implies the relation
\begin{equation}
\left( \frac{\dot{M}}{\dot{M}_\odot} \right)_\mathrm{W}
\stackrel{!}{=}
\frac{\dot{M}}{\dot{M}_\odot}
\frac{v_\infty}{v_{\infty,\odot}}
\ .
\label{mdotrel}
\end{equation}
If the terminal wind velocity is different from the solar value, then
the self-consistently determined mass loss rate, $\dot{M}$, is
different from the value, $\dot{M}_\mathrm{W}$, derived by
\citet{2002ApJ...574..412W}, since the latter is based on the
assumption of a unique solar-like wind velocity.
A comparison between theoretical and empirical stellar mass loss rates
is shown in Fig.\ \ref{maslos.fig} for a $1\,{\rm M_\odot}$ star
subject to different wind scenarios.
Note that, following Eq.\ (\ref{cwrpwood}), the relative CWRPs on the
abscissa are equivalent to the empirical mass loss rates per surface
area of \citet{2002ApJ...574..412W}.
For thermally driven winds of slow magnetic rotators, there is an
almost one-to-one correspondence between self-consistent and observed
mass loss rates, with the possible exception of very hot stellar
winds.
In contrast, in the regime of fast magnetic rotators, the terminal
velocities of magneto-centrifugally driven winds are faster than in the
solar case (see Fig.\ \ref{fmtv.fig}), and the self-consistent mass
loss rates thus smaller than those determined following the approach of
\citeauthor{2002ApJ...574..412W}.
In all but one of our wind scenarios the theoretical mass loss rates do
not exceed about ten times the solar value.
Only if the wind ram pressure scales exclusively with the wind density
are theoretical and empirical mass loss rates similar throughout.
In terms of this scenario, all stars are slow magnetic rotators with
solar-like winds, whose CWRPs scale with the mass flux.
The results for lower-mass stars are qualitatively the same, with
stellar mass loss rates (per surface area) not exceeding about ten
times the solar value either.
We conjecture that disregarding the importance of the (terminal) wind
velocity on the CWRP may lead to over-estimations of the mass loss
rates of rapidly rotating stars.

\section{Discussion}
\label{disc}
The aim of our investigation was to develop a picture of the wind ram
pressures and mass loss rates of cool main-sequence stars.
Rather than analysing individual stars in detail, we have used the set
of observationally determined values to constrain possible wind
scenarios.
In view of large observational uncertainties and the small and
heterogeneous sample of stars, we did not attempt to fit theoretical
and empirical values, but used the latter as guidance only.
The power-law ansatz for the dependence of the thermal and magnetic
wind parameters on the stellar rotation rate is motivated and supported
by previous investigations.
Nevertheless, with highest currently observed CWRPs occurring at
moderate rotation rates, it is not possible to find a consistent
agreement between a theoretical wind scenario and \emph{all} empirical
values.
This problem raises the question whether the increase of CWRPs is
characterised by the group of moderately rotating K dwarfs or by the
group of rapidly rotating targets.

\subsection{The K dwarf puzzle}
The high CWRPs of the K dwarfs \object{$\epsilon$ Eri}, \object{70 Oph}, 
and \object{36 Oph} cannot be explained in the framework of solar-like
magnetised winds.
The temperatures, densities, and heating rates required to raise the
values are inconsistent with the X-ray observations of
\citet{2006ApJ...643..444W}.
Strong magnetic fields could account for their CWRPs, but the observed
magnetic flux densities of these K dwarfs, albeit being unexpectedly
high, are insufficient.
On the observational side, inaccurate assumptions about the ambient ISM
or the difficult fitting procedure of simulated and observed
astrospheric absorption profiles may cause overestimations of the
empirical mass loss rates.
On the theoretical side, our model may miss out on additional wind
acceleration or energy transfer mechanisms, which could cause higher
terminal velocities.
However, we expect any effect capable of increasing the CWRP up to 80
times the solar value to render stellar winds non-solar and to imprint
discernible signatures on the coronal X-ray properties of the star.
Based on the available results, we consider the high CWRPs of the group
of K dwarfs to be peculiar and uncharacteristic for the winds of cool
main-sequence stars.

\subsection{CWRP of cool main-sequence stars}
Assuming that the increase of the CWRP of main-sequence stars is
characterised by the rapidly rotating targets \object{$\xi$ Boo} and
\object{EV Lac}, we argue in favour of stellar winds whose thermal and
magnetic properties increase moderately with the stellar rotation rate
like, for example, the wind scenario with $n_T= 0.1, n_n= 0.6$, and
$n_\Phi= 1$.
This set of parameters yields CWRPs in good agreement with the values
of the slow and rapidly rotating stars, and is based on previous,
independent investigations on the increase of thermal and magnetic
coronal properties \citep[e.g.][]{2003ApJ...599..516I,
2003SSRv..108..577F}.
However, the large uncertainties of the empirical mass loss rates also
allow for wind scenarios with parameters covering some range around
these reference values to be in agreement with the empirical
constraints.

Our theoretical results indicate that for a given rotation rate and
similar wind conditions, the CWRPs of cool stars increase toward later
spectral types.
The increase is due to the shrinking surface area; the mass loss rates
of lower mass stars are actually predicted to decrease.
The increase of the CWRP depends to some extent on our approximation of
stellar radii, $R\propto M^{0.8}$, on the lower main-sequence, and we
thus expect it to be diluted by the dependence on the actual radii and
mass loss rates of individual stars.
Since over the mass range of cool stars the difference is smaller than
the observational uncertainties, a structuring of CWRPs according to
stellar spectral type will be hardly discernible.
The situation worsens when the analysis is carried out in terms of the
coronal X-ray flux instead of the rotation, since the statistical
character of an rotation-activity relationship as well as intrinsic
(e.g.\ cyclic) variations of the X-ray emission add to the scatter of
empirical values.
In particular in the regime of fast magnetic rotators, we expect large
scatter of observed CWRPs around the mean values, caused by different
inclinations between stellar rotation axes and lines-of-sight.
The collimation of open magnetic field lines toward the stellar
rotation axis and non-uniform surface distributions of magnetic flux
make the wind ram pressure of rapidly rotating stars intrinsically
latitude-dependent \citep[e.g.][]{2000A&A...356..989T,
2005A&A...440..411H}.
Since astrospheres represent the impact of stellar winds averaged over
longitude and long (possibly decadal) timescales, small-scale and
intermittent variations will be smeared out.
Yet the typical concentration of magnetic flux at polar latitudes
\citep{2002AN....323..309S} as well as the collimation of magnetic
field lines may cause a gradient in the wind ram pressure between
equatorial and polar latitudes.
A more detailed analysis is required to quantify the possible impact of
these latitude-dependent effects.
Unfortunately, the inclination of a star is seldom known, so that these
effects can hardly be verified observationally.
But it is advisable to take them into account as a possible source of
scatter in the regime of fast magnetic rotators.

In their analysis of the X-ray properties of the moderately rotating K
dwarfs, \citet{2006ApJ...643..444W} find a possible connection between
the strength of their winds and coronal abundance anomalies.
Our investigation confirms a dependence of the CWRP on the FIP effect.
The influence on the mean molecular weight is small and the associated
impact on the CWRP well below the observational accuracy, but the trend
confirms the finding of \citeauthor{2006ApJ...643..444W}, that a strong
FIP effect implies a weaker stellar wind.
Yet we caution that our polytropic, hydrodynamic ansatz does probably
not allow for an adequate analysis of this question, and that a
dedicated investigation is required to clarify this point.

\subsection{Stellar mass loss rates}
Disregarding the terminal wind velocity in the analysis of CWRPs can
cause misestimations of stellar mass loss rates.
For fast magnetic rotators, the self-consistently determined mass loss
rates are typically smaller than the values based on the presumption of
solar-like winds, and do not exceed about ten times the solar value,
which is in agreement with upper limits based on observations of dMe
stars \citep{1996ApJ...462L..91L, 1997A&A...319..578V}.

Since the terminal wind velocities of fast magnetic rotators are higher
than solar-like values, mass loss rates more than an order of magnitude
higher than the actual values may be deduced.
Unfortunately, the wind velocities of cool stars are observationally
not constrained.
But due to the additional magneto-centrifugal driving mechanism a
similarity with surface escape velocities, like in the case of thermal
winds of slow rotators, cannot be expected per ser.
Very fast terminal wind velocities though require magnetic fields of
several kilo-Gauss, high filling factors, and high dynamo efficiencies,
which are observationally not confirmed and in contrast with current
theories on dynamo operation/saturation and on the rotational evolution
of cool stars.
Since the angular momentum loss associated with high magnetic fluxes
brakes the stellar rotation too efficiently, the observations of
rapidly rotating zero-age-main-sequence stars would be difficult to
explain \citep[and below]{2005A&A...444..661H}.
We consider wind scenarios based on high dynamo efficiencies to be
marginal, albeit respective values have been previously suggested
\citep[though focusing on a different aspect of stellar
activity]{2003ApJ...590..493S}.
The predicted range of dynamo efficiencies is consistent with the
observationally determined, super-linear values of
\citet{2001ASPC..223..292S}.

\subsection{Comparison of energy fluxes}
Combining our power-law approximations for the stellar mass loss rates
with the rotation-activity relationship, Eq.\ (\ref{deffxom}), yields
$\dot{M}\propto F_X^{n_F/n_X}$.
With the values given in Tables \ref{cwrpfit} and \ref{rotactrels}, the
rate of increase, $n_F/n_X\sim 0.5$, is significantly smaller than the
value $1.34\pm 0.18$, suggested by \citet{2005ApJ...628L.143W}.
The reason for this difference is that we assume the increase of CWRPs
to be characterised through the lower values of the rapidly rotating
targets, whereas \citeauthor{2005ApJ...628L.143W} base their fit on the
high CWRP of the peculiar group of moderately rotating K dwarfs.

The transport of polytropic gas from the stellar surface to infinity 
requires the specific thermal energy (per unit mass) 
\begin{equation}
q
= 
\frac{ \left( \gamma - \Gamma \right) }{
 \left( \gamma - 1 \right) \left( \Gamma - 1 \right) }
\frac{\Re}{\mu} T_0
\ ,
\label{defq}
\end{equation}
which has to be provided by the star.
Thus, at the stellar surface the thermal energy flux along open
magnetic fields must be
\begin{equation}
F_\mathrm{W}
=
\frac{q \dot{M}}{4\pi R^2}
=
\frac{q \mathcal{P}}{v_\infty}
\ ,
\end{equation}
or, in the case of $1\,{\rm M_{\sun}}$ stars with rotation rates
$\lesssim 4.2\Omega_\odot$, approximately
\begin{equation}
F_\mathrm{W}
\approx 
3.23\cdot10^4
\left( \frac{\Omega}{\Omega_\odot} \right)^{1.04}
\left[ \mathrm{\frac{erg}{s\, cm^2}} \right]
\ ,
\end{equation}
assuming a mono-atomic gas with a ratio of specific heats $\gamma=
5/3$.
At solar rotation rate, our approximation is in agreement with the
empirical values for quite sun regions, $F_\mathrm{W}\lesssim 5\cdot
10^4\,{\rm erg\cdot s^{-1}\cdot cm^{-2}}$, but an order of magnitude
smaller than the value for coronal holes \citep{1977ARA&A..15..363W}.
We note that the wind parameters we have used to gauge our model at
earth orbit (cf.\ Sect.\ \ref{src}) are characteristic for the slow
solar wind, which is expected to originate from quite sun regions,
whereas coronal holes are harbouring the fast solar wind component.
In comparison with the increase of the coronal X-ray flux, $F_X\propto
\Omega^{n_X}$ with $n_x\gtrsim 2$, the thermal energy flux of stellar
winds is predicted to increases with a much lower rate.

The coronal X-ray flux quantifies the energy input into closed magnetic
field regions, whereas the wind energy flux quantifies the (thermal)
energy input into wind regions.
Like \citet{2002ApJ...574..412W}, we do not expect a direct connection
between the two energy fluxes.
Closed magnetic field regions are mainly heated by the dissipation of
magnetic energy through reconnection processes.
\citet{2003ApJ...598.1387P}, for example, finds a correlation between
stellar X-ray luminosities and magnetic flux, $L_X\propto
\Phi^{1.15}$.
A dominant magnetic heating of wind regions is unlikely, since
reconnecting field lines would hamper the acceleration of coherent
plasma motions to super-sonic/super-Alfv\'enic flow velocities.
Instead, the Poynting flux presents an additional energy source for the
outflowing plasma, with magnetic field lines transferring rotational
energy of the star into kinetic energy without being dissipated.
The different rates of increase of the X-ray flux and the wind energy
flux may be due to different dependencies of the heating mechanisms on
the stellar rotation rate.
Correlating the stellar mass loss rate or wind energy flux with the
coronal X-ray flux may eliminate the explicit dependence on the
rotation rate, but it is questionable whether there is a direct
physical relationship between the two quantities.

For the X-ray luminosity-magnetic flux relation of
\citeauthor{2003ApJ...598.1387P} to be consistent with empirical
rotation-activity relationships (cf.\ Table \ref{rotactrels}) requires
dynamo efficiencies $n_\Phi\sim 2$, which would yield CWRPs outside the
errorbars of the rapidly rotating targets.
Such high dynamo efficiencies have severe ramifications for the
rotational evolution of cool stars, since high magnetic flux densities
imply a strong magnetic braking.
For instance, the observed spin-down of main-sequence stars with time,
$\Omega\propto \sqrt{t}$ \citep{1972ApJ...171..565S}, is consistent
with a linear dynamo efficiency.
The inconsistency between rotation-activity relations, on one side, and
stellar spin-down timescales, on the other, may be caused by the former
being related to the closed magnetic flux and the latter to the open
magnetic flux, indicating a different dependence of the open and closed
magnetic field structures on the stellar rotation rate.

\subsection{Impact on stars and planets}
\begin{figure}
\includegraphics[width=\hsize]{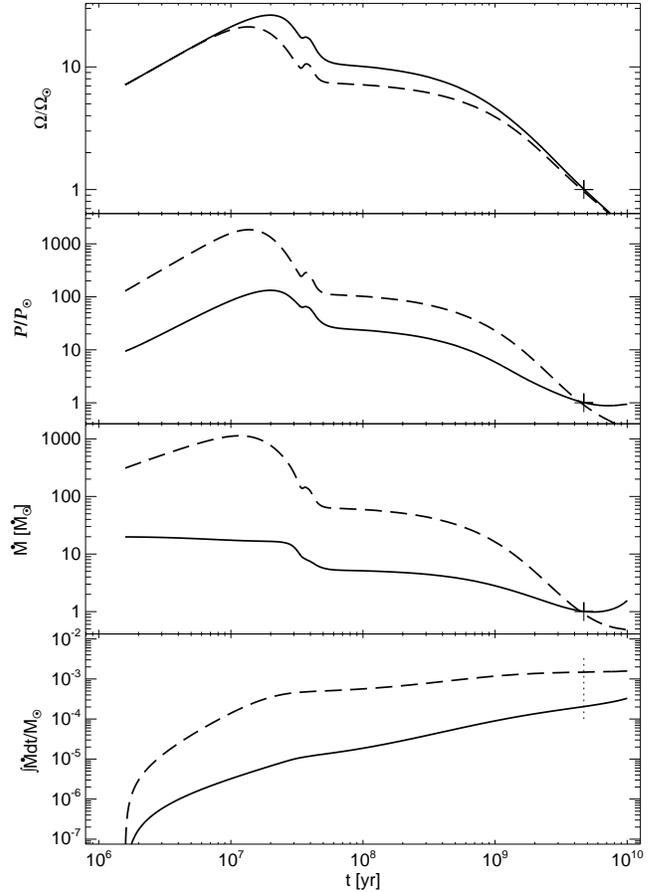}
\caption{Evolution of the stellar surface rotation rate, characteristic
wind ram pressure, mass loss rate, and cumulative mass loss (\emph{top}
to \emph{bottom}) of a $1\,{\rm M_{\sun}}$ star subject to wind
scenarios with a moderate increase of thermal and magnetic wind
parameters ($n_T= 0.1, n_n= 0.6, n_\Phi= 1$, \emph{solid lines}) and
with a strong increase of the wind density ($n_T= 0.1, n_n= 2, n_\Phi=
1$, \emph{dashed lines}), respectively.
The initial rotation rate is $\Omega_0= 2\cdot 10^{-6}\,{\rm s^{-1}}$,
and the internal coupling timescale $\tau_\mathrm{c}= 65\,{\rm Myr}$.
\emph{Crosses} mark the current state of the Sun.}
\label{rotevol.fig}
\end{figure}

\begin{figure}
\includegraphics[width=\hsize]{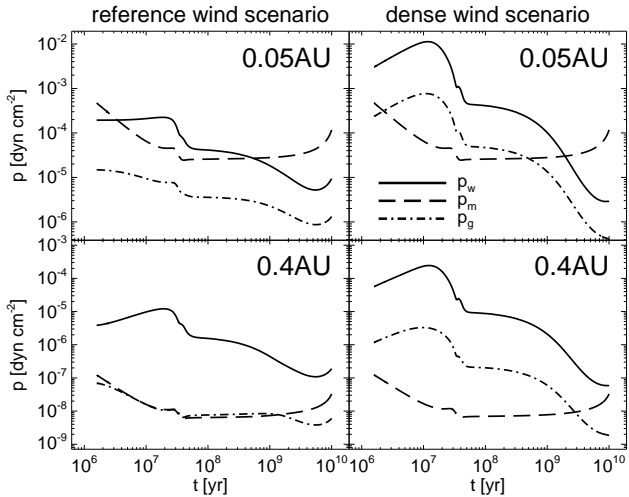}
\caption{Evolution of the wind ram pressure, $p_\mathrm{w}$, the
magnetic pressure, $p_\mathrm{m}$, and the thermal gas pressure,
$p_\mathrm{g}$, of magnetised winds at different distances inside the
equatorial plane.}
\label{prpr.fig}
\end{figure}

The increase of the mass loss rate with stellar rotation, $\dot{M}
\propto \Omega^{n_F}$, is according to our results ($n_F\lesssim 1$)
much weaker than predicted by \citet{2002ApJ...574..412W}: $n_F\approx
3.3$ [cf.\ combination of their Eqs.\ (1) and (3)].
For rapidly rotating stars, our mass loss rates are smaller than their
predictions, whereas for slowly rotating stars both predictions are
similar.
We illustrate the consequences of the different rates of increase by
comparing the effect of the reference and the dense wind scenarios on
the evolution of a $1\,{\rm M_{\sun}}$ star; the rotational evolution
model is described in \citet{2005A&A...444..661H}.
The latter scenario\footnote{Since for the dense wind scenario with
$n_n= 5$, analysed in Sect.\ \ref{resu}, the star loses about $65\%$ of
its initial mass within $20\,{\rm Myr}$, we consider instead the milder
case $n_n= 2$ to illustrate the principal differences.} is supposed to
reflect both the approach and result of \citet{2002ApJ...574..412W,
2005ApJ...628L.143W}, that stellar mass losses scale with the wind
density and increase strongly with the coronal X-ray flux/stellar
rotation rate.
For young stars, the CWRPs and mass loss rates based on the reference
wind scenario are several orders of magnitude smaller than in the case
of massive winds (Fig.\ \ref{rotevol.fig}).
The cumulated mass loss is $\sim 10^{-4}\,{\rm M_{\sun}}$, which
supports the conjecture of \citet{2002ApJ...574..412W} that the young
faint Sun-paradox cannot be explained through a higher-mass young Sun.
The impact of stellar winds on planetary atmospheres and magnetospheres
is most severe during the pre-main sequence phase (Fig.\
\ref{prpr.fig}).
The most promising targets for radio searches of extra-solar giant
planets are thus rapidly rotating pre-main sequence stars, which do not
only have strong winds but also high magnetic activity levels.
We note that this conjecture is based on the extrapolation of the
main-sequence wind scenarios to the pre-main sequence phase, since the
wind ram pressures of young stars are as yet observationally
unconstrained and thus hypothetical.

The magnetic energy flux at short-period orbits close to the star can
be larger than the wind ram pressure (Figs.\ \ref{prpr.fig} and
\ref{pprofiles.fig}), which may entail different interaction mechanisms
with planetary magnetospheres.
Still closer to the star, Hot Jupiters are expected to interact
directly with the coronal magnetic field.
Magnetic reconnections between coronal field structures and the
planetary magnetosphere alter the field topology and can initiate the
onset of flares and chromospheric brightenings
\citep{2004ApJ...602L..53I, 2005ApJ...622.1075S}.
That close to the stellar corona, wind ram pressure effects are
negligibly small \citep[e.g.][]{2006MNRAS.367L...1M}.

\subsection{Future issues}
It is essential to clarify the origin of the high CWRPs of the three K
dwarfs \object{$\epsilon$ Eri}, \object{36 Oph}, and \object{70 Oph},
because they play the decisive role in the determination of a
relationship between two fundamental stellar parameters.
Since their wind ram pressures cannot be explained in terms of a
solar-like wind model, we regard these stars as `non-representative'
and excluded them from the comparison of empirical and theoretical
values.
This approach may formally account for the lower CWRP of the more
rapidly rotating targets, but raises questions about inapplicabilities
of the polytropic-magnetised wind model or/and power-law ansatz for the
wind parameters.
\citeauthor{2002ApJ...574..412W}, in contrast, disregard the empirical
constraints set by the rapidly rotating targets and base their
relationship on the high values of the K dwarfs.
Their approach may formally account for the peculiar CWRPs, but raised
the question about different wind properties and coronal field
topologies beyond a certain stellar activity level.
The present stellar sample is insufficient to conclusively decide
between these two complementary approaches.
More astrospherical detections of (preferentially single) main-sequence
stars are highly desirable to solve this ambiguity.

If the high CWRP of the K dwarfs are observationally confirmed, then
this will raise crucial questions about our current understanding of
winds of cool stars, since the implicit requirements on the coronal
properties are drastically different from what we know and expect from
the Sun and solar-like stars.
Although previous observations indicate otherwise, the most likely
mechanism to account for such high wind ram pressure is an efficient
magneto-centrifugal driving of the wind due to strong magnetic fields.
Yet such high field strengths would raise the question how main
sequence stars with rotation rates not much different from the Sun can
produce and sustain such high magnetic flux densities over long times
without modifying the coronal X-ray signatures.

\section{Conclusions}
\label{conc}
Characteristic wind ram pressures and mass loss rates increase with the
wind temperature, wind density, and strength of open magnetic fields.
Albeit the observational data are as yet insufficient to conclusively
discriminate between different wind scenarios, we argue in favour of a
moderate increase of the thermal and magnetic wind properties with the
stellar rotation rate and, synonymously, the coronal X-ray flux.
Such a wind scenario does not account for the moderately rotating K
dwarfs \object{$\epsilon$ Eri}, \object{36 Oph}, and \object{70 Oph},
whose observed thermal and magnetic properties do not allow for a
consistent explanation of their high wind ram pressures in the
framework of a magnetised wind model.
We regard their high apparent mass loss rates as non-representative for
cool main-sequence stars and suggest to exclude them from quantitative
correlations between mass loss rates and rotation rates/coronal X-ray
fluxes until their role is observationally and theoretically
clarified.

The predicted rate of increase of mass loss rates is smaller than
suggested by \citet{2002ApJ...574..412W, 2005ApJ...628L.143W} and
depends on whether a star can be classified as a slow or a fast
magnetic rotator.
In the latter case, efficient magneto-centrifugal driving of outflows
entails terminal wind velocities considerably faster than the surface
escape velocity, which, if not taken properly into account, leads to an
overestimations of stellar mass loss rates.
The predicted mass loss rates of cool main-sequence stars do not exceed
(on average) about $10\,{\rm M_{\sun}}$.
Since the predicted stellar winds are weaker than previously suggested,
we expect less severe erosion of planetary atmospheres and lower
detectabilities of magnetospheric radio emissions originating from
extra-solar giant planets.
Considering the evolution of stellar mass loss rates and wind ram
pressure, we suggest that rapidly rotating pre-main sequence stars with
high magnetic activity levels are the most promising targets for
searches of planetary radio emission.

\begin{acknowledgements}
We thank the referee B.\ Wood for his very constructive comments which
helped to improve the paper.  
VH gratefully acknowledges financial support for this research through
a PPARC standard grand (PPA/G/S/2001/00144) and through a fellowship of
the Max-Planck-Society.
\end{acknowledgements}

\bibliographystyle{aa}
\bibliography{6486}

\end{document}